\documentclass[aps]{revtex4}
\usepackage{graphicx}

\usepackage{graphicx}

\newcommand{\be}{\begin{equation}}
\newcommand{\ee}{\end{equation}}
\newcommand{\bqn}{\begin{eqnarray}}
\newcommand{\eqn}{\end{eqnarray}}

\begin{document}

\title{Thermalization process in bare and dressed coordinate approaches}

\author{G. Flores-Hidalgo}
\email[E-mail address: ]{gflores@ift.unesp.br}
\affiliation{Instituto de Ci\^encias Exatas, Universidade Federal
de Itajub\'a, 37500-903, Itajub\'a, MG, Brazil}
\author{A. P. C. Malbouisson}
\email[E-mail address: ]{adolfo@cbpf.br} \affiliation{Centro
Brasileiro de Pesquisas F\'{\i}sicas/MCT, 22290-180, Rio de
Janeiro, RJ, Brazil}
\author{J. M. C. Malbouisson}
\email[E-mail address: ]{jmalboui@ufba.br} \affiliation{Instituto
de F\'{\i}sica, Universidade Federal da Bahia, 40.210-310,
Salvador (BA), Brazil}
\author{Y.W. Milla}
\email[E-mail address: ]{yonym@ift.unesp.br}
\affiliation{Instituto de F\'{\i}sica Teorica, Universidade
Estadual Paulista,  01405-900, S\~ao Paulo (SP), Brazil}
\author{A. E. Santana}
\email[E-mail address: ]{asantana@fis.unb.br}
\affiliation{Instituto de F{\'\i}sica, Universidade de
Bras{\'\i}lia, 70910-900, Bras{\'\i}lia-DF, Brasil}

\begin{abstract}
We consider a particle in the approximation of a harmonic
oscillator, coupled linearly to a   field modeling an environment.
The field is described by an infinite set of harmonic oscillators,
and the system (particle--field) is considered in a cavity at
thermal equilibrium. We employ the notions of bare and dressed
coordinates to study the time evolution of the occupation number.
With dressed coordinates no renormalization procedure is required,
leading directly to a finite result. In particular, for a large
time, the occupation number of the particle becomes independent of
its initial value. So we have a Markovian process, describing the
particle  thermalization  with the environment.
\end{abstract}

\date{\today}



\maketitle

A thermalization process occurs in some cases for a system of
material particles coupled to an environment, in
the sense that after an infinitely long time, the matter particles
 loose the memory of
their initial states. This study is, in general, not easy from a
theoretical point of view, due to the complex non--linear character
of the interactions between the matter particles and the environment. To get over these difficulties, linearized models  have
been adopted. An account on the subject of 
the evolution of quantum systems on general grounds can be found
 in~\cite{zurek,paz,ullersma,haake,caldeira,schramm}. Besides, the main analytical
method used
  to treat these systems at zero or finite
temperature is, except for a few
special cases, perturbation theory. In this framework, 
the perturbative  approach is carried out by means of the introduction of
{\it bare}, non--interacting  objects (fields, to which are associated bare quanta),
the interaction
being introduced order by order in powers of the coupling constant.

In spite of the remarkable achievements of the perturbative
methods, however,  there are situations where they cannot be employed, or are of little use.
 These cases have led  to attempts to
improve non-perturbative {\it analytical} methods,
 in particular,  where
strong effective couplings are involved. Among these trials  there
are  methods that perform resummations of perturbative series,
even if they are divergent, which amounts in some cases to extending 
the weak-coupling regime to a strong-coupling domain. One of these methods is the Borel resummation of perturbative series~\cite{rivasseau2,LeGuillou,Weniger,Jents,Cvetic,adolfoso}.  

In this paper we follow a different 
non-perturbative approach:  we investigate a simplified linear
version of a particle--field or particle--environment system,
where the particle, taken in the harmonic approximation, is
coupled to the reservoir, modeled by independent harmonic
oscillators~\cite{paz,ullersma,caldeira}. We will employ, in particular, \emph{dressed} states 
and {\it renormalized} coordinates. Using this method  non-perturbative treatments  can be considered for both weak and strong
 couplings.
 A linear model permits a better understanding of
 the need for non-perturbative analytical treatments
of coupled systems, which is the basic problem underlying the idea
of a \textit{dressed} quantum mechanical system. Of course, the
use of such an approach to a  realistic non-linear system is
an extremely hard task, while the linear model provides a good
compromise between physical reality and mathematical reliability.
The whole system is supposed to reside inside a spherical cavity
of radius $R$ in thermal equilibrium at temperature
$T=\beta^{-1}$. In other words, we consider the spatially
regularized theory (finite $R$) at finite temperature. The free
space case is obtained by suppressing the regulator,
($R\rightarrow\infty$). For a detailed comparison between this
procedure and the one considering an {\it a priori} unbounded
space, see~\cite{adolfo1}.
\section{The model}
%
%
%
Let us start by considering a particle approximated by a harmonic
oscillator, having \emph{bare} frequency $\omega _0$, linearly
coupled to a set of $N$ other harmonic oscillators, with
frequencies $\omega _k$, $k=1,2,\ldots ,N$. The Hamiltonian for
such a system is written in the form,
\begin{equation}
H=\frac 12\left[ p_0^2+\omega _0^2q_0^2+\sum_{k=1}^N\left(
p_k^2+\omega _k^2q_k^2\right) \right] - q_0\sum_{k=1}^Nc_kq_k,
\label{Hamiltoniana}
\end{equation}
leading to the following equations of motion,

\begin{eqnarray}
\ddot{q}_0+\omega_0^2q_0&=&\sum_{i=1}^Nc_{i}q_i(t)\label{eqmot291} \\
\ddot{q}_i+\omega_i^2q_i&=&c_{i}q_0(t)\label{eqmot292}.
\end{eqnarray}
In the limit $N\rightarrow
\infty $, we recover our case of the particle coupled to the
environment, after redefining divergent quantities, in a manner
analogous to mass renormalization  in field theories.
A Hamiltonian of the type (\ref{Hamiltoniana}) has been largely
used in the literature, in particular  to study the quantum
Brownian motion with the path-integral formalism~\cite{zurek,paz}. It has also
been employed  to
investigate the linear coupling of a particle to the scalar
potential~\cite{adolfo1,adolfo2,adolfo3,adolfo4,adolfo5}.

The Hamiltonian (\ref{Hamiltoniana}) is transformed to principal
axis by means of a point transformation, 
\begin{eqnarray}
q_\mu  &=&\sum_{r=0}^{N}t_\mu ^rQ_r\,,\,\,\,p_\mu =\sum_{r=0}^{N}t_\mu ^rP_r;\qquad   \nonumber \\
\mu  &=&(0,\{k\}),\qquad k=1,2,...,N;\;\;r=0,...N,  \label{transf}
\end{eqnarray}
performed by an orthonormal matrix $T=(t_\mu ^r)$. The subscripts
$\mu =0$ and $\mu =k$ refer respectively to the particle and the
harmonic modes of the reservoir and $r$ refers to the normal
modes. In terms of normal momenta and coordinates, the transformed
Hamiltonian  reads
\begin{equation}
H=\frac 12\sum_{r=0}^N(P_r^2+\Omega _r^2Q_r^2),  \label{diagonal}
\end{equation}
where the $\Omega _r$'s are the normal frequencies corresponding
to the collective \textit{stable} oscillation modes of the coupled
system. Using the coordinate transformation~(\ref{transf}) in the
equations of motion  and explicitly making use of the
normalization of the matrix $(t_{\mu}^{r})$, $\sum_{\mu =0}^N(t_\mu ^r)^2=1$,    
 we get
\begin{equation}
t_k^r=\frac{c_k}{\omega _k^2-\Omega _r^2}t_0^r\;,\;\;t_0^r=\left[
1+\sum_{k=1}^N\frac{c_k^2}{(\omega _k^2-\Omega _r^2)^2}\right] ^{-\frac 12},
\label{tkrg1}
\end{equation}
with the condition
\begin{equation}
\omega _0^2-\Omega _r^2=\sum_{k=1}^N\frac{c_k^2}{\omega _k^2-\Omega _r^2}.
\label{Nelson1}
\end{equation}

We take $c_k=\eta (\omega _k)^u$, where $\eta $ is a constant
independent of $k$. In this case the environment is classified
according to $u>1$, $u=1$, or $u<1$, respectively as
\textit{supraohmic}, \textit{ohmic} or \textit{subohmic}. This
terminology has been used in studies of the quantum Brownian
motion and of dissipative
systems~\cite{paz,ullersma,haake,caldeira,schramm}. For a subohmic
environment the sum in Eq.~(\ref{Nelson1}) is convergent in the
limit $N\rightarrow \infty $ and the frequency $\omega _0$ is well
defined. For ohmic and supraohmic environments, this sum  diverges for $N\rightarrow
\infty $. This makes the equation meaningless, unless  a
renormalization procedure is implemented. From now on we restrict
ourselves to an \textit{ohmic} system. In this case,
 Eq. (\ref {Nelson1}) is 
written in the form
\begin{equation}
\omega _0^2-\delta \omega ^2-\Omega _r^2=\eta ^2\Omega
_r^{2}\sum_{k=1}^N\frac 1{\omega _k^2-\Omega _r^2},
\label{Nelson2}
\end{equation}
where we have defined the counterterm
\begin{equation}
\delta \omega ^2=N\eta ^2.  \label{omegabarra1}
\end{equation}
%
%
%
There are $N+1$ solutions of $\Omega _r$, corresponding to the
$N+1$ normal collective modes. Let us for a moment suppress the
index $r$ of $\Omega _r^2$. If $\omega _0^2>\delta
\omega ^2$, all possible solutions for $\Omega ^2$ are positive,
physically meaning that the system oscillates harmonically in all
its modes. If $\omega _0^2<\delta \omega ^2$,
then  a single negative solution exists. In order to
prove this let us define  the function
\begin{equation}
I(\Omega ^2)=\omega _0^2-\delta \omega ^2-\Omega ^2-\eta ^2\Omega
^{2}\sum_{k=1}^N\frac 1{\omega _k^2-\Omega ^2},
\end{equation}
so that Eq.~(\ref{Nelson2}) becomes $I\left( \Omega ^2\right) =0$.
We find that

\[
 I(\Omega ^2) {\rightarrow} \infty  \,\,\, {\rm{as}} \,\,\, \Omega
^2\rightarrow -\infty \,\,\,{\rm{and}} \,\,\,I(0)=\omega _0^2-\delta \omega ^2<0, 
\]
in the interval $(-\infty , 0]$. As $I\left( \Omega ^2\right) $ is
a monotonically decreasing function in this interval, we conclude
that $I\left( \Omega ^2\right) =0$ has a single negative solution
in this case. This means that there is a mode whose amplitude
grows or decays exponentially, so that no stationary
configuration is allowed. Nevertheless, it should be remarked that
in a different context, it is precisely this runaway solution that
is related to the existence of a bound state in the
Lee--Friedrichs model. This solution is considered in  the framework of a model to describe
qualitatively the existence of bound states in particle physics~\cite{Likhoded}.

Considering the situation where all normal modes are harmonic,
which corresponds to the first case above, $\omega _0^2>\delta
\omega ^2$, we define the \textit{renormalized} frequency
\begin{equation}
\bar{\omega}^2=\omega _0^2-\delta \omega ^2=\lim _{N \rightarrow
\infty }(\omega_{0}^2 - N\eta^2), \label{omegabarra}
\end{equation}
in terms of which Eq.~(\ref{Nelson2})  in the limit
$N\rightarrow \infty $ becomes,
\begin{equation}
\bar{\omega}^2-\Omega ^2=\eta ^2\sum_{k=1}^\infty \frac{\Omega ^{2}}{%
\omega _k^2-\Omega ^2}.
\label{Nelson3}
\end{equation}
In this limit, the above procedure is exactly the
analog of the mass renormalization in quantum field theory: the
addition of a counterterm $-\delta \omega ^2q_0^2$ allows one to
compensate the infinity of $\omega _0^2$ in such a way as to leave
a finite, physically meaninful renormalized frequency
$\bar{\omega}$. This simple renormalization scheme has been
 introduced earlier~\cite{Thirring}. Unless explicitly stated, 
the limit $N\rightarrow \infty $ is
understood in the following. 

Let us define a constant $g$, with dimension of frequency, by 
\begin{equation}
g=\frac{\eta ^2}{2\Delta \omega},
\label{g}
\end{equation}
where $\Delta \omega =\pi c/R$.
The environment  frequencies $\omega _k$ are given by,
\begin{equation}
\omega _k=k\frac{\pi c}R,\;\;\;\;k=1,2,\ldots \;,
\label{discreto}
\end{equation}
where $R$ is the radius of the cavity that contains the whole
system. Then, using the identity
\begin{equation}
\sum_{k=1}^\infty \frac 1{k^2-u^2}=\frac 12\left[ \frac 1{u^2}-\frac \pi
u\cot \left( \pi u\right) \right] ,  \label{id4}
\end{equation}
Eq. (\ref{Nelson3}) can be written in a closed form:

\begin{equation}
\cot \left( \frac{R\Omega }c\right) =\frac \Omega {\pi g}+\frac
c{R\Omega }\left( 1-\frac{R\bar{\omega}^2}{\pi gc}\right).
\label{eigenfrequencies1}
\end{equation}
The solutions of the above equation with respect to $\Omega $ give
the spectrum of eigenfrequencies $\Omega _r$ corresponding to the
collective normal modes.

In terms of the physically meaningful quantities $\Omega _r$ and
$\bar{\omega}$, the transformation matrix elements turning the
particle--field system to the principal axis are obtained. They
are 
\begin{eqnarray}
t_0^r &=&\frac{\eta \Omega _r}{\sqrt{(\Omega _r^2-\bar{\omega}^2)^2+\frac{%
\eta ^2}2(3\Omega _r^2-\bar{\omega}^2)+\pi ^2g^2\Omega _r^2}},  \nonumber \\
t_k^r &=&\frac{\eta \omega _k}{\omega _k^2-\Omega _r^2}t_0^r.
\label{t0r2}
\end{eqnarray}
These matrix elements 
play a central role in the quantities describing the system.

\section{The thermalization process in bare coordinates}

We now consider the thermalization problem using bare coordinates.
For the model described by Eq. (\ref{Hamiltoniana}) this problem
was  addressed in an alternative way in~\cite{gabrielrudnei}
with the canonical Liouville-von Neumann formalism. We consider
the initial state described by the density operator,

\begin{equation} \rho(t=0)=\rho_0\otimes\rho_\beta\;,
\label{q1}\end{equation}
where $\rho_0$ is the density operator of the particle, that
in principle can be in a pure or in a mixed state and
$\rho_\beta$ is the density operator of the thermal bath, at
a temperature $\beta ^{-1}$, that is,

\begin{equation} \rho_\beta=
Z_\beta^{-1}\exp\left[-\beta\sum_{k=1}^{\infty}\omega_k\left(a_k^\dag a_k+\frac{1}{2}
\right)\right]\;,
\label{q2}
\end{equation}
with $Z_\beta=\prod_{k=1}^Nz_\beta^k$ being the partition function
of the reservoir, and

\begin{equation} z_\beta^k= {\rm
Tr}_k\left[e^{-\beta\omega_k\left(a_k^\dag a_k+1/2\right)}\right]
=\frac{1}{2\sinh\left(\frac{\beta_k\omega_k}{2}\right)}\; ;
\label{pfun}\end{equation}
The creation and
annihilation operators given by

\begin{eqnarray}a_\mu&=&\sqrt{\frac{\bar{\omega}_\mu}{2}}q_\mu+
\frac{i}{\sqrt{2\bar{\omega}_\mu}}p_\mu
\label{o1}\\
a_\mu^\dag&=& \sqrt{\frac{\bar{\omega}_\mu}{2}}q_\mu-
\frac{i}{\sqrt{2\bar{\omega}_\mu}}p_\mu\;, \label{0o1}
\end{eqnarray}   where  $\bar{\omega}_\mu =
(\bar{\omega},\omega_k)$. The thermalization problem is 
addressed by investigating the time evolution of the state
$\rho(t)$.

The thermalization problem concerns the time evolution of the initial state
 to thermal equilibrium. The subsystem corresponding to
the particle oscillator is described by an arbitrary density
operator $\rho_0$. As we will show, the expectation value of
the number operator corresponding to particles will evolve in
time to a value that is independent of the initial density
operator $\rho_0$, the dependence will be exclusively on the
mixed density operator corresponding to the thermal bath.

Our aim is to obtain expressions for the time evolution of the
expectation values for the occupation number and in particular
for the one corresponding to  particles. We will
solve the problem in the framework of the Heisenberg picture. It
is to be understood that when a quantity appears without the time
argument it means that such quantity is evaluated at $t=0$. The
Heisenberg equation of motion for the  annihilation operator
$a_\mu(t)$ is given by
\begin{equation}
 \frac{\partial}{\partial
t}a_\mu(t)=i\left[\hat{H},a_\mu(t)\right]\;.
\label{q5}
\end{equation}
Due to the linear character of our problem, this equation is 
solved by writing  $a_\mu(t)$ as
\begin{equation} a_\mu(t)=\sum_{\nu=0}^{\infty}\left(\dot{B}_{\mu\nu}(t)\hat{q}_\nu+
B_{\mu\nu}(t)\hat{p}_\nu\right)\;, \label{q6}\end{equation}
where all the time dependence is in the {\it c -number} functions
$B_{\mu\nu}(t)$. Then,   Eq.~(\ref{q5}) reduces to  the following
coupled equations for $B_{\mu\nu}(t)$:

\begin{equation} \ddot{B}_{\mu 0}(t)+\bar{\omega}^2B_{\mu 0}(t)-\sum_{k=1}^{\infty}
\eta \omega_k B_{\mu k}(t)=0, \label{q7}\end{equation}

\begin{equation} \ddot{B}_{\mu k}(t)+\omega_k^2B_{\mu k}(t)-B_{\mu
0}(t)\sum_{k=1}^{\infty} \eta \omega_k=0 . \label{q8}\end{equation}

These equations are formally identical to the classical
equations of motion,   Eqs.~(\ref{eqmot291})  and
(\ref{eqmot292}), for the bare coordinates $q_\mu$. Then we 
decouple Eqs.~(\ref{q7}) and (\ref{q8}) with the same matrix
$\{t_\mu^r\}$ that diagonalizes the Hamiltonian
Eq.~(\ref{Hamiltoniana}). In an analogous manner,
we write $B_{\mu\nu}(t)$ as

\begin{equation} B_{\mu\nu}(t)=\sum_{r=0}^{\infty}t_\nu^rC_\mu^r(t), \label{ec7}
\end{equation}
such that from  Eqs.~(\ref{q7}) and (\ref{q8}), we obtain the
following equations for the normal-axis functions $C_\mu^r(t)$,

\begin{equation}  \ddot{C}_\mu^r(t)+\Omega_r^2C_\mu^r(t)=0,\label{ex1}
\end{equation}
 which gives the solution  

\[
C_\mu^r(t)=a_\mu^re^{i \Omega_r t}+ b_\mu^re^{-i\Omega_r t} .
\]
Then substituting this  expression into Eq.~(\ref{ec7}) we find
\begin{equation} B_{\mu\nu}(t)=\sum_{r=0}^{\infty}t_\nu^r\left(a_{\mu}^re^{i\Omega_r
t}+ b_{\mu}^re^{-i\Omega_r t}\right). \label{q9}
\end{equation}
The time independent coefficients $a_{\mu}^r$, $b_{\mu}^r$ are
determined by the initial conditions at $t=0$ for $B_{\mu\nu}(t)$
and $\dot{B}_{\mu\nu}(t)$. From Eqs.~(\ref{o1}) and (\ref{q6}) we
find that these initial conditions are given by

\begin{eqnarray}B_{\mu\nu}&=&\frac{i\delta_{\mu\nu}}{\sqrt{2\bar{\omega}_\mu}},\nonumber\\
\dot{B}_{\mu\nu}&=&\sqrt{\frac{\bar{\omega}_\mu}{2}}\delta_{\mu\nu}\;.
\label{o2} \end{eqnarray}%
Using these equations, we obtain for $a_\mu^r$ and $b_\mu^r$,

\begin{eqnarray}a_\mu^r&=&\frac{it_\mu^r}{\sqrt{8 \bar{\omega}_{\mu}}}
\left(1-\frac{\bar{\omega}_\mu}{\Omega_r}\right),
\label{o3a}\\
b_\mu^r&=&\frac{it_\mu^r}{\sqrt{8\bar{\omega}_{\mu}}}\left(1+\frac{\omega_\mu}{\Omega_r}\right)\;.
\label{o3b} \end{eqnarray}%
We write $a_\mu (t)$ and $a_\mu^\dag (t)$ in terms of $a_\mu $ and $a_\mu^\dag $ using Eqs.~(\ref{o1}), (\ref{0o1}) and (\ref{q6}),  
\begin{eqnarray}
a_\mu(t)&=&\sum_{\nu=0}^{\infty}\left(\alpha_{\mu\nu}(t)\hat{a}_\nu
+\beta_{\mu\nu}(t)\hat{a}^\dag_\nu\right)\, ,
\label{0o4} \\
a_\mu^\dag(t)&=&\sum_{\nu=0}^{\infty}\left(\beta^\ast_{\mu\nu}(t)\hat{a}_\nu
+\alpha_{\mu\nu}^\ast(t)\hat{a}^\dag_\nu\right)\, , \label{o4}
\end{eqnarray}
where $\alpha_{\mu\nu}(t)$ and $\beta_{\mu\nu}(t)$ are the  Bogoliubov
coefficients given by,

\begin{equation} 
\alpha_{\mu\nu}(t)=\frac{1}{\sqrt{2\omega_\nu}}\dot{B}_{\mu\nu}(t)-
i\sqrt{\frac{\omega_\nu}{2}}B_{\mu\nu}(t)
\label{o5}
\end{equation}
and
\begin{equation} \beta_{\mu\nu}(t)=\frac{1}{\sqrt{2\omega_\nu}}\dot{B}_{\mu\nu}(t)+
i\sqrt{\frac{\omega_\nu}{2}}B_{\mu\nu}(t)\;.
\label{o6}
\end{equation}

Using the definition of $B_{\mu\nu}(t)$ we get 
\begin{eqnarray}
\alpha_{\mu\nu}(t)&=&\sum_{r=0}^{\infty}\sqrt{\frac{\omega_\nu}{\omega_\mu}}
\frac{t_\mu^rt_\nu^r}{4\Omega_r}
\left\{\frac{\Omega_r}{\omega_\nu} \left[(\omega_\mu-\Omega_r)
e^{i\Omega_r t}+(\omega_\mu+\Omega_r) e^{-i\Omega_r t}\right]\right. \nonumber \\
&&\left. +\left[(\Omega_r-\omega_\mu)e^{i\Omega_r t} +
(\Omega_r+\omega_\mu)e^{-i\Omega_r t}\right]\right\}, \label{o7}
\end{eqnarray}%
and
\begin{eqnarray}\beta_{\mu\nu}(t)&=&\sum_{r=0}^{\infty}\sqrt{\frac{\omega_\nu}{\omega_\mu}}
\frac{t_\mu^rt_\nu^r}{4\Omega_r}\left\{\frac{\Omega_r}{\omega_\nu}
\left[(\omega_\mu-\Omega_r)e^{i\Omega_r t}+(\omega_\mu+\Omega_r)
e^{-i\Omega_r t}\right] \right. \nonumber \\
&&\left. -\left[(\Omega_r-\omega_\mu)e^{i\Omega_r t}+
(\Omega_r+\omega_\mu)e^{-i\Omega_r t}\right]\right\}\;. \label{o8}
\end{eqnarray}
Now we study the time evolution of $n_\mu(t)$, the expectation
value of the number operator
$N_\mu(t)=a^\dag_\mu(t)a_\mu(t)$, that is,
\begin{equation} n_\mu(t) ={\rm Tr}\left[a^\dag_\mu(t)a_\mu(t)
\rho_0\otimes\rho_\beta\right]\;.
\label{o9}\end{equation}
Using the basis
$|n_0,n_1,n_2,...n_N\rangle$  we obtain,

\begin{equation} n_\mu(t)=\sum_{\nu=0}^{\infty}\left[|\alpha_{\mu\nu}(t)|^2+|\beta_{\mu\nu}(t)|^2\right]
n_\nu+\sum_{\nu=0}^{\infty}|\beta_{\mu\nu}(t)|^2\;,
\label{o10}
\end{equation}
where

\begin{equation} n_0=\sum_{n=0}^{\infty}n\langle n|\rho_0|n\rangle
\label{o11}
\end{equation}
is the  expectation value of the number operator
corresponding to the particle  and the set $\{n_k\}$
stands for the  thermal expectation values corresponding to the
thermal bath oscillators, given by the Bose-Einstein distribution,

\begin{equation} n_k=\frac{1}{e^{\beta\omega_k}-1}\;.
\label{o12}
\end{equation}

In Eq. (\ref{o10}) there appears a term that does not depend 
on the temperature of the thermal bath. This term has its origin
in the instability of the initial bare vacuum state,
$|0,0,...,0\rangle$. To see this, we  compute the expectation
value of the time dependent number operator
$N_{\mu}(t)=a_\mu^\dag(t)a_\mu(t)$ in this
 vacuum state. Thus all the terms containing operators
different from the identity give a zero contribution. The only
term, that gives a non-zero contribution comes from the normal
ordering and 
is just the last one in Eq.~(\ref{o10}). This term leads to the creation of excited states (particles, in a
field theoretical language) from the initial unstable bare vacuum
state.

We are interested in evaluating the expectation value of
the number operator corresponding to  particles. Thus taking
$\mu=0$ in Eq.~(\ref{o10}) and using Eq.~(\ref{o12}), we obtain

\begin{eqnarray}n_0(t)&=&\left[|\alpha_{00}(t)|^2+|\beta_{00}(t)|^2\right]n_0
+\sum_{k=1}^{\infty}\left[|\alpha_{0k}(t)|^2+|\beta_{0k}(t)|^2\right]
\frac{1}{e^{\beta\omega_k}-1} \nonumber \\
&&+|\beta_{00}(t)|^2+\sum_{k=1}^{\infty}|\beta_{0k}(t)|^2\; ,
\label{ec9a} \end{eqnarray}%
where the coefficients of this expression are~\cite{gabrielrudnei}, 
\begin{eqnarray} 
\alpha_{00}(t)&=&\frac{e^{-\pi g t/2}}{16\bar{\omega}\kappa}
\left[\left(2\bar{\omega}+2\kappa-i\pi g\right)^2 e^{-i\kappa t}-
\left(2\bar{\omega}-2\kappa-i\pi g\right)^2 e^{i\kappa
t}\right]\;, \label{ec10}\\
\beta_{00}(t)&=&\frac{\pi ge^{-\pi g t/2}}{8\bar{\omega}\kappa}
\left[\left(\pi g+2i\kappa\right)e^{-i\kappa t}- \left(\pi
g-2i\kappa\right)e^{i\kappa t}\right]\;, \label{ec11}\\
\alpha_{0k}(t)&=&\sqrt{\frac{\omega_k}{2\bar{\omega}}}
\frac{(\bar{\omega}+\omega_k)\sqrt{g\Delta \omega}\,e^{-i\omega_k
t}} {\left(\omega_k^2-\bar{\omega}^2+i\pi g\omega_k\right)}+
\sqrt{\frac{\omega_k}{\bar{\omega}}}\frac{\sqrt{2
g\Delta\omega}} {4\kappa} \nonumber\\
&&\times \left[\frac{(2\kappa+2\bar{\omega}-i\pi g)}
{(2\kappa-2\omega_k-i\pi g)}e^{-i\kappa t}
+\frac{(2\bar{\omega}-2\kappa-i\pi g)}{(2\kappa+2\omega_k+i\pi g)}
e^{i\kappa t}\right]e^{-\pi g t/2} \label{ec12}
 \end{eqnarray}%

and

\begin{eqnarray}\beta_{0k}(t)&=&\sqrt{\frac{\omega_k}{2\bar{\omega}}}
\frac{(\omega_k-\bar{\omega})\sqrt{g\Delta \omega}\,e^{i\omega_k
t}} {\left(\omega_k^2-\bar{\omega}^2-i\pi
g\omega_k\right)}-\sqrt{\frac{\omega_k}{\bar{\omega}}}
\frac{\sqrt{2g\Delta\omega}}{4\kappa} \nonumber\\
& & \times \left[\frac{(2\bar{\omega}+2\kappa-i\pi
g)}{(2\kappa+2\omega_k-i\pi g)} e^{-i\kappa
t}+\frac{(2\bar{\omega}-2\kappa-i\pi g)} {(2\kappa-2\omega_k+i\pi
g)} e^{i\kappa t}\right]e^{-\pi g t/2}\;, \label{ec13} \end{eqnarray}%
such that

\begin{equation} \kappa=\sqrt{\bar{\omega}^2-\pi^2g^2/4}.
\label{kappa}\end{equation}

The parameter $\kappa$  measures
the intensity of the interaction: if $\kappa^2>>0$, $i.e$ 
 $g<<2\bar{\omega}/\pi$,  we are in the {\it weak}
coupling regime. On the contrary if $\kappa^2<<0$, $i.e$ 
$g>>2\bar{\omega}/\pi$, the system is in the {\it strong} coupling
regime. 
Here we will restrict ourselves to the weak coupling regime. This
case includes the important class of electromagnetic
interactions,  $g=\alpha
\bar{\omega}$, with $\alpha$ being the fine structure constant
$\alpha =1/137$~\cite{adolfo2}.

In the continuum limit $\Delta\omega\to 0$, sums over $k$
become integrations over a continuous variable $\omega$ and
 we obtain for $n_0(t)$,

\begin{eqnarray}n_0(t)&=&\frac{e^{-\pi gt}}{\bar{\omega}^2\kappa^2}
\left[\bar{\omega}^4+\frac{\pi^2g^2}{8}\left(2\bar{\omega}^2-\pi^2g^2\right)\cos(2\kappa
t) -\frac{\pi^3g^3\kappa}{4}\sin(2\kappa t)\right]n_0 \nonumber \\
&&+\frac{\pi^2g^2e^{-\pi gt}}{16\bar{\omega}^2\kappa^2}
\left[2\bar{\omega}^2+\left(2\bar{\omega}^2-\pi^2g^2\right)\cos(2\kappa
t)- 2\pi g\kappa\sin(2\kappa t)\right]\nonumber\\
& &+\frac{g}{\bar{\omega}} \int_0^\infty d\omega
\left[\frac{F(\omega,\bar{\omega},g,t)}{\left(e^{\beta\omega}-1\right)}+G(\omega,\bar{\omega},g,t)\right]
, \label{ec14} \end{eqnarray}%
where

\begin{eqnarray}F(\omega,\bar{\omega},g,t)&=&
\frac{\omega(\omega^2+\bar{\omega}^2)}
{\left[(\omega^2-\bar{\omega}^2)^2+\pi^2g^2\omega^2\right]}
\left\{1+ \frac{e^{-\pi gt}}{4\kappa^2} [4\bar{\omega}^2-\pi^2g^2\cos(2\kappa t)   \right. \nonumber \\
&&  -2\pi
g\kappa\frac{(\omega^2-\bar{\omega}^2)}{(\omega^2+\bar{\omega}^2)}
\sin(2\kappa t) ] -\frac{e^{-\pi gt/2}}{\kappa}
[2\kappa\cos(\omega
t)\cos(\kappa t)     \nonumber\\
& &   +
\frac{4\omega\bar{\omega}^2}{(\omega^2+\bar{\omega}^2)}\sin(\omega
t)\sin(\kappa t)
 \left. -\pi
g\frac{(\omega^2-\bar{\omega}^2)}{(\omega^2+\bar{\omega}^2)}
\cos(\omega t)\sin(\kappa t) ] \right\} \nonumber \\
\label{ec15}
\end{eqnarray}%
and

\begin{eqnarray}G(\omega,\bar{\omega},g,t)&=&
\frac{\omega(\omega-\bar{\omega})^2}
{\left[(\omega^2-\bar{\omega}^2)^2+\pi^2g^2\omega^2\right]}\left\{1+\frac{e^{-\pi
gt}}{4\kappa^2}
\left[4\bar{\omega}^2+\frac{2\pi^2g^2\bar{\omega}\omega}
{(\omega-\bar{\omega})^2} \right. \right. \nonumber\\
& &\left. -
\pi^2g^2\frac{(\omega^2+\bar{\omega}^2)}{(\omega-\bar{\omega})^2}
\cos(2\kappa t)-2\pi g
\kappa\frac{(\omega+\bar{\omega})}{(\omega-\bar{\omega})}
\sin(2\kappa t)\right]\nonumber\\
& & -\frac{e^{-\pi gt/2}}{\kappa}  [2\kappa\cos(\omega
t)\cos(\kappa t) -2\bar{\omega}\sin(\omega t)\sin(\kappa t)
\nonumber \\
&& \left. -\pi g
\frac{(\omega+\bar{\omega})}{(\omega-\bar{\omega})}\cos(\omega
t)\sin(\kappa t)] \right\}\;. \label{ec16} \end{eqnarray}%

It is to be noted that the second and the third lines in Eq.~(\ref{ec14}) are
independent of the initial distributions. Also the integral in the third line of 
$G(\omega,\bar{\omega},g,t) $ is  logarithmically divergent. We
can understand the origin of this divergence in the following way:
suppose that initially, in the absence of the linear interaction,
we prepare the system in its ground state, that is, at $t=0$ we
have $|0,0,...,0\rangle$. Then, we can compute, in the Heisenberg
picture, the time evolution for the expectation value of the
number operator corresponding to the particle, that is $\langle
0,0,...,0|\hat{a}_0^\dag(t)\hat{a}_0(t)|0,0,...,0\rangle$.
Taking $\mu=0$ we obtain,

\begin{equation} \langle 0,0,...,0|\hat{a}_0^\dag(t)\hat{a}_0(t)|0,0,...,0\rangle=
|\beta_{00}(t)|^2+\sum_{k=1}^{\infty}|\beta_{0k}(t)|^2\;,
\label{ec17}
\end{equation}
which in the continuum limit gives the second line of
Eq.~(\ref{ec14}). Then, the origin of the  divergence
appearing in Eq.~(\ref{ec14}) is interpreted as the
excitations  produced from the unstable bare (vacuum) ground state, as a
response to the linear interaction. 

As we are interested in the thermal behavior of $n_0(t)$ only, the
second line and the   term $G(\omega,\bar{\omega},g,t)$ in the
third line of Eq.~(\ref{ec14}) can be neglected. This is a  renormalization procedure. Thus we write the following {\it renormalized} expectation value
 for the particle number
operator,

\begin{equation} \bar{n}_0(t)=K(\bar{\omega},g,t)n_0
+\frac{g}{\bar{\omega}}\int_0^\infty d\omega
\frac{F(\omega,\bar{\omega},g,t)}{\left(e^{\beta\omega}-1\right)}
\label{ec18}
\end{equation}
where

\begin{equation} K(\bar{\omega},g,t)= \frac{e^{-\pi gt}}{\bar{\omega}^2\kappa^2}
\left[\bar{\omega}^4+\frac{\pi^2g^2}{8}\left(2\bar{\omega}^2-\pi^2g^2\right)\cos(2\kappa t)
-\frac{\pi^3g^3\kappa}{4}\sin(2\kappa t)\right]\;.
\label{ec19}
\end{equation}

In the limit $t\to\infty$, ${\bar{n}}_{0}(t)$ has a well defined
value, that is, the system reaches a final equilibrium state.
Also, since $K(\bar{\omega},g,t\to\infty)\to 0$, this final
equilibrium state is independent of $n_0$. The equilibrium
expectation value of the number operator corresponding to the
particle is independent of its initial value, and the only
dependence is on the initial distribution of the thermal bath,
that is, the particle
 thermalizes with the environment.  
 Before the interaction enters into
play for $t< 0$, $n(t< 0)=n_0$, then we have that
$K(\omega,\bar{\omega},g,t<0)=1$. Taking $t=0$ in Eq. (\ref{ec19})
we obtain $K(\omega,\bar{\omega},g,t=0)=
\bar{\omega}^2/\kappa^2+\pi^2g^2(2\bar{\omega}^2-\pi^2g^2)/(8\bar{\omega}^2\kappa^2)$.
Thus $K(\omega,\bar{\omega},g,t)$
is a discontinuous function of $t$; the discontinuity appearing
just at $t=0$. From the physical standpoint this discontinuity can be viewed as a
response to the sudden onset of the interaction between 
particles and the environment.

\begin{figure}[t]
\includegraphics[{height=8.0cm,width=9cm,angle=360}]{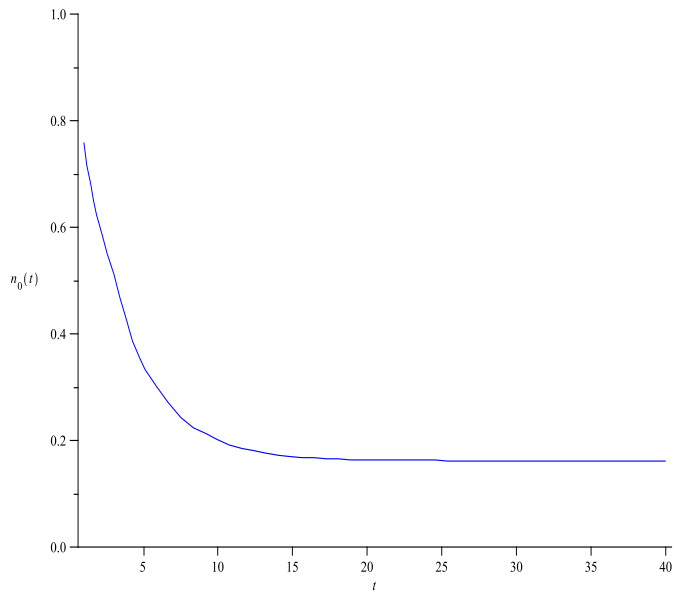}
\caption{Time behavior for $\bar{n}_0(t)$ given
by Eq. (\ref{ec18}) for ($t>1$), $n_0=1$, $\bar{\omega}=1$,
$\beta=2$ and $g=0.1$}
\label{figuraXXIX1}
\end{figure}

Although the integral in Eq. (\ref{ec18}) can not be computed
analytically, we can perform  numerical calculations, for example
in Fig. {\ref{figuraXXIX1}} we display the time behavior for $n_0=1$,
$\bar{\omega}=1$, $\beta=2$ and $g=0.1$; ($t>1$).
In the next section we develop an alternative approach based on
the notion of dressed particles. We will find that, in this new realm,  no renormalization is needed.
%
%
%
%

%

\section{Dressed coordinates and dressed states}

Let us start with the eigenstates of our system, $\left|
n_0,n_1,n_2...\right\rangle $, represented by the normalized
eigenfunctions in terms of the normal coordinates $\{Q_r\}$,
\begin{equation}
\phi _{n_0n_1n_2...}(Q,t)=\prod_s\left[ \sqrt{\frac{2^{n_s}}{n_s!}}%
H_{n_s}\left( \sqrt{\frac{\Omega _s}\hbar }Q_s\right) \right]
~\Gamma _0e^{-i\sum_sn_s\Omega _st},  \label{autofuncoes}
\end{equation}
where $H_{n_s}$ stands for the $n_s$-th Hermite polynomial and
$\Gamma _0$ is the normalized vacuum eigenfunction, 
\begin{equation}
\Gamma_{0}={\cal{N}}e^{-\frac{1}{2}\sum_{r=0}^{\infty}\Omega_{r}^{2}Q_{r}^{2}}
\label{eigenvacuo}
\end{equation}.

We introduce \textit{dressed} or \textit{renormalized} coordinates $%
q_0^{\prime }$ and $\{q_i^{\prime }\}$ for, respectively, the \textit{dressed%
} particle and the \textit{dressed} field, defined by, 
\begin{equation}
\sqrt{\bar{\omega}_\mu }q_\mu ^{\prime }=\sum_rt_\mu
^r\sqrt{\Omega _r}Q_r, \label{qvestidas1}
\end{equation}
valid for arbitrary $R$ and where $\bar{\omega}_\mu
=\{\bar{\omega},\;\omega _i\}$.
In terms of dressed coordinates, we define for a fixed
instant, $t=0$, \textit {dressed} states, $\left| \kappa _0,\kappa
_1,\kappa _2...\right\rangle $ by means of the complete
orthonormal set of functions 
\begin{equation}
\psi _{\kappa _0\kappa _1...}(q^{\prime })=\prod_\mu \left[ \sqrt{\frac{%
2^{\kappa _\mu }}{\kappa _\mu !}}H_{\kappa _\mu }\left( \sqrt{\frac{\bar{%
\omega}_\mu }\hbar }q_\mu ^{\prime }\right) \right] \Gamma _0,
\label{ortovestidas1}
\end{equation}
where $q_\mu ^{\prime }=\left\{ q_0^{\prime },\,q_i^{\prime }\right\} $, $%
\bar{\omega}_\mu =\{\bar{\omega},\,\omega _i\}$. Notice that the ground state $%
\Gamma _0$ in the above equation is the same as in
Eq.(\ref{autofuncoes}). The invariance of the ground state is due
to our definition of dressed coordinates given by Eq.
(\ref{qvestidas1}). 
Each function $\psi _{\kappa _0\kappa
_1...}(q^{\prime })$ describes a state in which the dressed
oscillator $q_\mu ^{\prime }$ is in its $\kappa _\mu $-th excited
state.

It is worthwhile to note that our renormalized coordinates are
new objects, different from both the bare
coordinates, $q$, and the normal coordinates $Q$. In particular, the
renormalized coordinates and dressed states, although both are 
collective objects, should not be confused with the
normal coordinates $Q$, and the eigenstates Eq.~(\ref{autofuncoes}).
While the eigenstates $\phi$ are
stable, the dressed states $\psi$ are all unstable, except for the ground
 state obtained by setting $\{\kappa _\mu =0\}$ in Eq.
(\ref{ortovestidas1}). The idea is that the dressed states are  physically
meaningful states. This can be seen as an analog of
the wave-function renormalization in quantum field theory, which
justifies the denomination of \textit{renormalized} to the new
coordinates $q^{\prime }$. Thus, the dressed state given by Eq.
(\ref{ortovestidas1}) describes the particle in its $\kappa _0$-th
excited level and each mode $k$ of the cavity in the $\kappa
_k-th$ excited level.  It should be noticed that the introduction
of the renormalized coordinates guarantees the stability of the dressed vacuum state, since by definition it is identical to
the ground state of the system. The fact that the definition given by
Eq. (\ref{qvestidas1}) assures this requirement can be easily seen
by replacing Eq. (\ref{qvestidas1}) in Eq. (\ref{ortovestidas1}). We
obtain $\Gamma _0(q^{\prime })\propto \Gamma _0(Q)$, which shows
that the dressed vacuum state given by Eq. (\ref {ortovestidas1})
is the same ground state of the interacting Hamiltonian given by
Eq.~(\ref{diagonal}).

The necessity of introducing renormalized coordinates can be
understood by considering what would happen if we write Eq.
(\ref{ortovestidas1}) in terms of the bare coordinates $q_\mu $.
In the absence of interaction, the bare states are stable since
they are eigenfuntions of the free Hamiltonian. But when we consider the
interaction they all become unstable. The excited states
are unstable, since we know this from experiment. On the other hand, we also know from experiment
that the particle in its ground state is stable, in contradiction
with what our simplified model for the system describes in terms
of the bare coordinates. So, if we wish to have a nonperturbative
approach in terms of our simplified model something should be
modified in order to remedy this problem. The solution is just the
introduction of the renormalized coordinates $q_\mu ^{\prime }$ as
the physically meaningful ones.

In terms of  bare coordinates, the dressed coordinates are
expressed as
\begin{equation}
q_\mu ^{\prime }=\sum_\nu \alpha _{\mu \nu }q_\nu ,
\label{qvestidas3}
\end{equation}
where
\begin{equation}
\alpha _{\mu \nu }=\frac 1{\sqrt{\bar{\omega}_\mu }}\sum_rt_\mu ^rt_\nu ^r%
\sqrt{\Omega _r}.  \label{qvestidas4}
\end{equation}
If we consider an arbitrarily large cavity ($R\rightarrow \infty
$), the dressed coordinates reduce
to 
\begin{eqnarray}
q_0^{\prime } &=&A_{00}(\bar{\omega},g)q_0,  \label{q0'q0} \\
q_i^{\prime } &=&q_i,  \label{qi'qi}
\end{eqnarray}
with $A_{00}(\bar{\omega},g)$ given by,
\begin{equation}
A_{00}(\bar{\omega},g)=\frac 1{\sqrt{\bar{\omega}}
}\int_0^\infty \frac{2g\Omega ^2\sqrt{\Omega }d\Omega }{(\Omega ^2-\bar{%
\omega}^2)^2+\pi ^2g^2\Omega ^2} . \label{alfa00}
\end{equation}
In other words, in the limit $R\rightarrow \infty $, the particle
is still dressed by the field, while for the field there remain bare
modes.

Let us consider a particular dressed state $\left| \Gamma
_{1}^{\mu}(0)\right\rangle $, represented by the wavefunction
$\psi _{00\cdots 1(\mu )0\cdots }(q^{\prime })$. It describes the
configuration in which only the dressed oscillator $q_\mu
^{\prime }$ is in the {\it first} excited level. Then  the
following expression for its time evolution  is
valid~\cite{adolfo1}:
\begin{eqnarray}
\left|\Gamma _{1}^{\mu} (t)\right\rangle  &=&\sum_\nu f^{\mu \nu
}(t)\left| \Gamma _{1}^{\nu} (0)\right\rangle\,  \nonumber \\
f^{\mu \nu }(t) &=&\sum_st_\mu ^st_\nu ^se^{-i\Omega _st}.
\label{ortovestidas5}
\end{eqnarray}
Moreover we find that
\begin{equation}
\sum _{\nu}\left|f^{\mu \nu}(t)\right|^{2}=1\,.
\label{probabilidade}
\end{equation}
Then 
the coefficients $%
f^{\mu \nu }(t)$ are simply interpreted as   probability
amplitudes.

 In approaching the thermalization process in this
framework, we have to write the initial
physical state in terms of dressed coordinates, or equivalently in
terms of dressed annihilation and creation operators
$a_\mu'$ and $a_\mu'^{\dag}$ instead of $a_\mu$
and $a_\mu^{\dag}$. This means that the 
initial dressed  density operator corresponding to the thermal bath is
given by

\begin{equation} \rho_\beta=Z_\beta ^{-1}
\exp\left[-\beta\sum_{k=1}^{\infty}\omega_k\left(a_k'^{\dag}a_k'+\frac{1}{2}
\right)\right], \label{ec6}\end{equation}
where we define 

\begin{eqnarray}a_\mu'&=&\sqrt{\frac{\bar{\omega}_\mu}{2}}q_\mu'+
\frac{i}{\sqrt{2\bar{\omega}_\mu}}p_\mu'
\label{a1}\\
a_\mu'^\dag&=&
\sqrt{\frac{\bar{\omega}_\mu}{2}}q_\mu'-
\frac{i}{\sqrt{2\bar{\omega}_\mu}}p_\mu'\;. \label{q4} \end{eqnarray}%
Now  we analyze the time evolution of dressed
coordinates.
\section{Thermal behavior for a cavity of arbitrary size with dressed coordinates}

The solution for the time-dependent  annihilation and creation
dressed operators follows similar steps as for the bare
operators. The time evolution of the
annihilation operator is given by,
\begin{equation}
\frac{d}{dt}a_\mu'(t) =
i\left[\hat{H},a_\mu'(t)\right]\; \label{v1}
\end{equation}
and a similar equation for $a_\mu'^\dag(t)$. We solve
this equation with the initial condition at $t=0$,
\begin{equation}
a_\mu'(0)=\sqrt{\frac{\omega_\mu}{2}}q_\mu'+
\frac{i}{\sqrt{2\omega_\mu}}p_\mu'\;,
 \label{v2}
\end{equation}
which, in terms of bare coordinates, becomes
\begin{equation}
a_\mu'(0)=\sum_{r,\nu=0}^N\left(\sqrt{\frac{\Omega_r}{2}}
t_\mu^rt_\nu^r\hat{q}_\nu
+\frac{it_\mu^rt_\nu^r}{\sqrt{2\Omega_r}}\hat{p}_\nu\right)\;.
\label{q10}
\end{equation}
We assume a solution for $a_\mu'(t)$ of the type
\begin{equation}
a_\mu'(t)=\sum_{\nu=0}^{\infty}\left(\dot{B}_{\mu\nu}'(t)\hat{q}_\nu+
B_{\mu\nu}'(t)\hat{p}_\nu\right)\;. \label{v3}
\end{equation}
Using Eq.(\ref{Hamiltoniana}) we find, 
\begin{equation}
B_{\mu\nu}'(t)=\sum_{r=0}^{\infty}t_\nu^r\left(a_{\mu}'^re^{i\Omega_r t}+
b_{\mu}'^re^{-i\Omega_r t}\right)\;. \label{B1}
\end{equation}
In the present case the time independent coefficients are
different from those in the bare coordinate approach,
Eq.~(\ref{q9}). The initial conditions for $B_{\mu\nu}'(t)$ and
$\dot{B}_{\mu\nu}'(t)$ are obtained by setting $t=0$ in
Eq.~(\ref{v3}) and comparing with Eq.~(\ref{q10}); Then 
\begin{eqnarray}
B_{\mu\nu}'(0)&=&i\sum_{r=0}^{\infty}\frac{t_\mu^rt_\nu^r}{\sqrt{2\Omega_r}}\;,
\label{q11}\\
\dot{B}_{\mu\nu}'(0)&=&\sum_{r=0}^{\infty}\sqrt{\frac{\Omega_r}{2}}t_\mu^rt_\nu^r\;.
\label{q12} \end{eqnarray}
Using these initial conditions and the orthonormality of the
matrix $\{t_\mu^r\}$ we obtain $a_\mu'^r=0$,
$b_\mu'^r=it_\mu^r/\sqrt{2\Omega_r}$. Replacing these values for
$a_\mu'^r$ and $b_\mu'^r$ in Eq.~(\ref{B1}) we get
\begin{equation}
B_{\mu\nu}'(t)=i\sum_{r=0}^{\infty}\frac{t_\mu^rt_\nu^r}
{\sqrt{2\Omega_r}}e^{-i\Omega_r t}\;. \label{q13}
\end{equation}
 We have
\begin{eqnarray} a_\mu'(t)&=&\sum_{r,\nu=0}^Nt_\mu^rt_\nu^r
\left(\sqrt{\frac{\Omega_r}{2}}\hat{q}_\nu
+\frac{i}{\sqrt{2\Omega_r}}\hat{p}_\nu\right)e^{-i\Omega_r t}\nonumber\\
&=&\sum_{r,\nu=0}^Nt_\mu^rt_\nu^r\left(\sqrt{\frac{\omega_\nu}{2}}\hat{q}_\nu'+
\frac{i}{\sqrt{2\omega_\nu}}\hat{p}_\nu'\right)e^{-i\Omega_r t}
=\sum_{\nu=0}^{\infty}f_{\mu\nu}(t)\hat{a}_\nu'\;, \label{q14}
\end{eqnarray}
where
\begin{equation}
f_{\mu\nu}(t)=\sum_{r=0}^{\infty} t_\mu^rt_\nu^re^{-i\Omega_r t}\;.
\label{ex2}
\end{equation}

For the occupation number $n_\mu'(t)=\langle
a_\mu'^\dag(t)a_\mu'(t)\rangle$ we get 
\begin{equation} n_{\mu}'(t)={\rm Tr}\left(a_\mu'^\dag(t)a_\mu'(t)
\rho_0'\otimes\rho_\beta'\right)\;. \label{q15}
\end{equation} where $\rho_0'$ is the density operator for the
dressed particle and $\rho_\beta'$ is the density operator
for the thermal bath, which coincides with the corresponding
operator for the bare thermal bath if the system is in free space
(in the sense of an arbitrarily large
cavity)\cite{adolfo1,adolfo2}.

To evaluate $n'_\mu(t)$  we choose the basis
$|n_0,n_1,...,n_N\rangle=\prod_{\mu=0}^{\infty}|n_\mu\rangle$, where
$|n_\mu\rangle$ are the eigenvectors of the number operators
$a'^{\dag }_{\mu}a'_\mu$. From Eq. (\ref{q14}) we get
\begin{eqnarray}
a_\mu'^\dag(t)a_\mu'(t)&=&\sum_{\nu,\rho=0}^{\infty}f_{\mu\rho}^\ast(t)
f_{\mu\nu}(t)\hat{a}_\rho'^\dag\hat{a}_\nu'
\nonumber\\
&=&\sum_{\nu=0}^{\infty}|f_{\mu\nu}(t)|^2 \hat{a}'^\dag_\nu \hat{a}_\nu'+
\sum_{\nu\neq\rho}f_{\mu\rho}^\ast(t)f_{\mu\nu}(t)
\hat{a}'^\dag_\nu\hat{a}_\rho'\;. \label{q16} \end{eqnarray}
In the basis $|n_0,n_1,n_2,\cdots \rangle$  we obtain,

\begin{equation}
n_\mu'(t)=|f_{\mu 0}(t)|^2n_0'+\sum_{k=1}^{\infty}|f_{\mu k}(t)|^2n_k'\;,
\label{q17}
\end{equation}
where $n_0'$ and $n_k'$ are the expectation values of the initial number
operators, respectively,  for the dressed particle and  dressed
bath modes. We assume that,  dressed field modes obey a Bose-Einstein
distribution. This can be justified by remembering that in the
free space limit, $R \rightarrow \infty$, dressed field modes
are identical to the bare ones, according to Eqs.~(\ref{q0'q0})
and (\ref{qi'qi}). Now, no term independent of the temperature
appears in the thermal bath. This should be expected since  the
dressed vacuum is stable, particle production from the vacuum is
not possible. Setting $\mu=0$ in Eq.~(\ref{q17}) we obtain the
time evolution for the ocupation number  of the particle,

\begin{equation}
n_{0}'(t)=|f_{0 0}(t)|^2n_0'+\sum_{k=1}^{\infty}|f_{0 k}(t)|^2n_k'\;.
\label{q18}
\end{equation}

\section{The limit of arbitrarily large cavity: unbounded space}

In a large cavity (free space) we must compute the quantities
$f_{00}(t)$ and $f_{0k}(t)$ in the
continuum limit  to study the
time evolution of the ocupation number for the particle.Remember 
that in Eqs.~(\ref{t0r2}), $\omega_k=k\pi c/R$, $k=1,2,...$ and
$\eta= \sqrt{2g\Delta \omega}$, with $\Delta \omega=(\omega_{i+1}-
\omega_i)=\pi c/R$. When $R\to\infty$,  we have
$\Delta \omega\to 0$ and $\Delta \Omega\to 0$ and then, the sum in Eq.~(\ref{ex2}) becomes an integral. To calculate the quantities
$f_{\mu\nu}(t)$ we first note that, in the continuum limit,
Eq.~(\ref{t0r2}) becomes 
\begin{eqnarray}
t_0^r &\rightarrow & t_0^{\Omega}\sqrt{\Delta \Omega } \equiv
\lim_{\Delta \Omega \rightarrow 0}\frac{\Omega \sqrt{2g\Delta
\Omega}}
{\sqrt{(\Omega ^2-\bar{\omega}^2)^2+\pi ^2g^2\Omega ^2}},  \label{t0r0(1)} \\
t_k^r &\rightarrow &\frac{\omega \sqrt{2g\Delta \omega}}{\omega
^2-\Omega ^2}t_0^{\Omega}\sqrt{\Delta \Omega }. \label{t0r2(1)}
\end{eqnarray}
 In the following, we
suppress the labels in the frequencies, since they are continuous
quantities.

We start by defining a function $W(z)$,
\begin{equation}
W(z)=z^2-\bar{\omega}^2+\sum_{k=1}^{\infty}\frac{\eta^2z^2}
{\omega_{k}^{2}-z^2}\;. \label{cc5}
\end{equation}
We find that the
$\Omega$'s are the roots of $W(z)$. Using  $\eta^2=2g\Delta
\omega$, we have in the continuum limit, 
\begin{equation}
W(z)=z^2-\bar{\omega}^2+2gz^2\int_{0}^{\infty}\frac{d\omega}
{\omega^2-z^2}\;. \label{cc6}
\end{equation}
For complex values of $z$ the above integral is well defined and
is evaluated by using Cauchy theorem, to be 
\begin{equation}
W(z)=\left\{\begin{array}{c}
z^2+ig\pi z-\bar{\omega}^2,~{\rm{Im}}(z)>0\\
z^2-ig\pi z-\bar{\omega}^2,~{\rm{Im}}<0\;.
\end{array}\right.
\label{cc7}
\end{equation}

We now compute $f_{00}(t)=\sum_{r=0}^{\infty} (t_0^r)^2
e^{-i\Omega_r t}$ which, in the continuum limit, is given by
\begin{equation}
f_{00}(t)=\int_{0}^{\infty}(t_0^{\Omega})^2e^{-i\Omega t}\,d\Omega
\;. \label{cc8}
\end{equation}
We find
that,
\begin{equation}
(t_0^{\Omega})^2=\frac{1}{W(\Omega)}\;, \label{cc9}
\end{equation}
and since the $\Omega$'s are the roots of $W(z)$, we  write
Eq.~(\ref{cc8}) as
\begin{equation}
f_{00}(t)=\frac{1}{i\pi}\oint_C\frac{dz e^{-izt}}{W(z)}\;,
\label{cc10}
\end{equation}
where $C$ is a counterclockwise contour in the $z$-plane that
encircles the real positive roots of $W(z)$. Choosing a contour
infinitesimally close to the positive real axis, that is
$z=\alpha-i\epsilon$ below it and $z=\alpha+i\epsilon$ above it
with $\alpha>0$ and $\epsilon\to 0^+$, we obtain
\begin{equation}
f_{00}(t)=\frac{1}{i\pi}\int_{0}^{\infty}d\alpha \alpha
e^{-i\alpha t}\left[\frac{1}{W(\alpha-i\epsilon)}-\frac{1}
{W(\alpha+i\epsilon)}\right]\;. \label{cc11}
\end{equation}
In the limit $\epsilon\to 0^+$, Eq.~(\ref{cc7}) gives $W(\alpha\pm
i\epsilon)=\alpha^2-\bar{\omega}^2\pm ig\pi\alpha$ which leads to
\begin{equation}
f_{00}(t) = C_1(t;\bar{\omega},g) + i S_1(t;\bar{\omega},g),
\label{f00}
\end{equation}
where
\begin{eqnarray}
C_1(t;\bar{\omega},g) & = & 2g \int_{0}^{\infty} d\alpha
\frac{\alpha^2 \cos(\alpha t)}{(\alpha^2-\bar{\omega}^2)^2 +
\pi^2 g^2\alpha^2} , \\
S_1(t;\bar{\omega},g) & = & - 2g \int_{0}^{\infty} d\alpha
\frac{\alpha^2 \sin(\alpha t)}{(\alpha^2-\bar{\omega}^2)^2 + \pi^2
g^2\alpha^2} . \label{S1}
\end{eqnarray}
Notice that $C_1(t=0;\bar{\omega},g)=1$ and
$S_1(t=0;\bar{\omega},g)=0$, so that $f_{00}(t=0) = 1$ as expected
from the orthonormality of the matrix $(t_\mu^r)$. The real part
of $f_{00}(t)$ is calculated using the residue theorem. For
$\kappa^2 = \bar{\omega}^2 - \pi^2 g^2 /4 > 0$, which includes the
weak coupling regime, one finds
\begin{equation}
C_1(t;\bar{\omega},g) = e^{-\pi g t/2}\left[\cos(\kappa t)
-\frac{\pi g}{2\kappa} \sin(\kappa t)\right] \;\;\;\; (\kappa^2 >
0). \label{C1}
\end{equation}
Although $S_1(t;\bar{\omega},g)$ cannot be
analytically evaluated for all $t$, however for long times, i.e. $t \gg
1/\bar{\omega}$, we have
\begin{equation}
S_1(t;\bar{\omega},g) \approx
\frac{4g}{\bar{\omega}^4t^3}\;\;\;\;\;(t\gg \frac 1{\bar{\omega}
}). \label{J1}
\end{equation}
Thus, we get for large $t$
\begin{equation}
|f _{00}(t)|^2 \approx e^{-\pi gt} \left[ \cos(\kappa t)
-\frac{\pi g}{2\kappa} \sin(\kappa t) \right]^2 + \frac{16
g^2}{\bar{\omega}^{8}t^6}. \label{efe002}
\end{equation}

Next we compute the quantity $f_{0k}(t)=\sum_{r=0}^{\infty} t_0^r
t_k^r e^{-i\Omega_r t}$ in the continuum limit. It is 
\begin{equation}
f_{0\omega}(t)=\eta\omega\int_{0}^{\infty}\frac{(t_0^{\Omega})^2
e^{-i\Omega t}d\Omega
}{(\omega^2-\Omega^2)}=\frac{\eta\omega}{i\pi}\oint_C\frac{ze^{-izt}}
{(\omega^2-z^2)W(z)}\;, \label{cc14}
\end{equation}
where  $\eta =\sqrt{2g\Delta \omega}$. Taking the same
contour as that used to calculate $f_{00}(t)$, we obtain
\begin{equation}
f_{0\omega}(t)=-\frac{\eta\omega}{i\pi}\int_0^{\infty}d\alpha
\left[\frac{\alpha e^{-i\alpha t}}{W(\alpha-i\epsilon)
[(\alpha-i\epsilon)^2-\omega^2]}-\frac{\alpha e^{-i\alpha t}}
{W(\alpha+i\epsilon)[(\alpha+i\epsilon)^2-\omega^2]}\right]\;.
\label{cc16}
\end{equation}
Thus, taking $\epsilon\to 0^+$ 
$f_{0\omega}(t)$ is written as
\begin{equation}
f_{0\omega}(t) = \omega \sqrt{\Delta\omega} \left[
C_2(\omega,t;\bar{\omega},g) + i S_2(\omega,t;\bar{\omega},g)
\right] \;, \label{f0w}
\end{equation}
where
\begin{eqnarray}
C_2(\omega,t;\bar{\omega},g) & = & (2g)^{\frac{3}{2}}
\int_{0}^{\infty} d\alpha \frac{\alpha^2 \cos(\alpha t)}{(\omega^2
- \alpha^2)\left[(\alpha^2-\bar{\omega}^2)^2+
\pi^2 g^2\alpha^2\right]} , \\
S_2(\omega,t;\bar{\omega},g) & = & - (2g)^{\frac{3}{2}}
\int_{0}^{\infty} d\alpha \frac{\alpha^2 \sin(\alpha t)}{(\omega^2
- \alpha^2) \left[ (\alpha^2-\bar{\omega}^2)^2 + \pi^2 g^2\alpha^2
\right]} . \label{S2}
\end{eqnarray}
Notice that the integrals defining the functions $C_2$ and $S_2$
are actually Cauchy principal values.

\begin{figure}[t]
\includegraphics[{height=8.0cm,width=9cm,angle=360}]{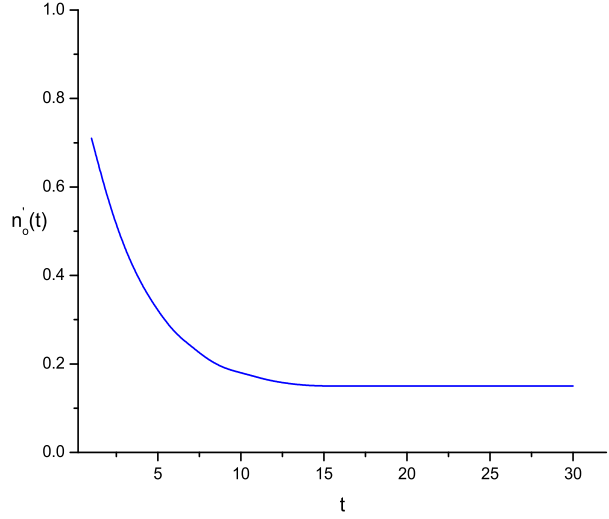}
\caption{Time behavior for $n_0'(t)$ given by Eq. (\ref{number}),
for ($t>1$), $n_0=1$, $\bar{\omega}=1$, $\beta=2$ and $g=0.1$}
\label{figuraXXIX2}
\end{figure}

The function $C_2$ is calculated analytically using Cauchy
theorem; we find
\begin{eqnarray}
C_2(\omega,t;\bar{\omega},g) & =& \sqrt{2g} \left[ e^{-\pi g t/2}
\left\{ \frac{\omega^2 - \bar{\omega}^2}{(\omega^2 -
\bar{\omega}^2)^2 + \pi^2 g^2 \omega^2}\, \cos{\kappa t}\right.\right.\nonumber \\
&&\left. -
\frac{\pi g}{2\kappa} \frac{\omega^2 + \bar{\omega}^2}{(\omega^2 -
\bar{\omega}^2)^2 + \pi^2 g^2 \omega^2}\, \sin{\kappa t} \right\}
\nonumber \\
 & & \left.  + \, \frac{\pi g \omega}{(\omega^2 -
\bar{\omega}^2)^2 + \pi^2 g^2 \omega^2}\, \sin{\omega t} \right] .
\label{C2}
\end{eqnarray}
The function $S_2$ cannot be  evaluated analytically 
for all $t$, it has to be calculated numerically. For long times,
 we have
\begin{equation}
S_2(t;\bar{\omega},g) \approx \frac{4\sqrt{2}g\sqrt{g}}{\omega^2
\bar{\omega}^4 t^3}\;\;\;\;\;(t\gg \frac 1{\bar{\omega} }).
\label{J2}
\end{equation}
In the continuum limit, we get the average of the particle ocupation number, 
\begin{equation}
n^{\prime}_{0}(t) = \left[ C_1^2(t;\bar{\omega},g) +
S_1^2(t;\bar{\omega},g) \right] n^{\prime}_0 + \int_{0}^{\infty}
d\omega \,\omega^2 \left[ C_2^2(\omega,t;\bar{\omega},g) +
S_2^2(\omega,t;\bar{\omega},g) \right]n^{\prime}(\omega) \,,
\label{number}
\end{equation}
where $n^{\prime}(\omega)=1/(e^{\beta \omega} - 1)$ is the density
of occupation of the environment modes, the functions $C_1$ and
$C_2$ are given by Eqs.~(\ref{C1}) and (\ref{C2}) while the
functions $S_1$ and $S_2$ are given by the integrals Eqs.~(\ref{S1})
and (\ref{S2}), respectively.
In Fig. \ref{figuraXXIX2} we display the behavior in time for $n_0=1$,
$\bar{\omega}=1$, $\beta=2$ and $g=0.1$; ($t>1$).

The important point, that is seen from Fig.~\ref{figuraXXIX1} and Fig.~\ref{figuraXXIX2} is that, for long times, both the bare and dressed ocupation numbers of the particle approach smoothly to the same asymptotic value. Moreover this value is the one expected on physical grounds, obtained from the Bose distribution at the final equilibrium temperature. In fact, taking $\beta=2$ and $\bar{\omega}=1$, as used in the plots, we get
$$ n_{\infty}(\bar{\omega})=1/(e^{\beta \bar{\omega}} - 1)=0.156$$
Therefore both methods, and in particular our dressed state formalism describes very precisely the thermalization process.

\section{Final remarks}
We have considered a linearized version of a particle–-environment
system and we have carried out a non--perturbative  treatment
of the thermalization process. We have adopted the point of view of renouncing to an approach very close to the real behavior of a 
nonlinear system, to study instead a linear model.
As a counterpart, an exact solution has been possible. This realises a good compromise between physical reality and mathematical reliability. 
We have presented an ohmic quantum system consisting of a particle, in the
larger sense of a material body, an atom or a Brownian particle coupled
to an environment modelled by non-interacting oscillators. We have used 
the formalism of  dressed states to perform a non-perturbative
study of the time evolution of the system, contained in a cavity or in free space. 
Distinctly to what happens in the bare coordinate approach, in the
dressed coordinate approach no renormalization procedure is
needed. Our renormalized
coordinates 
contain in themselves the renormalization aspects. 
As far as the thermalization process is concerned from a physical viewpoint, 
both bare and dressed approaches are in agreement with what we expect for this process. For long times, all
the information about the particle occupation numbers depends only on the environment. 
Both curves in Fig.~\ref{figuraXXIX1} and Fig.~\ref{figuraXXIX2} approach steadly to an asymptotic value of the bare and dressed ocupation numbers of the particle, which is the physically expected one at the given temperature. 

\section*{Acknowledgements}
We are specially grateful to F.C. Khanna for fruitful discussions and the Theoretical Physics Institute, University of Alberta, for kind hospitality during the summer 2008. Our thanks also to P.J. Pompeia and R.R. Cuzinatto for help in numerical calculations. This work received partial financial support from  CNPq/MCT and FAPERJ.

\end{document}